\documentclass[twocolumn]{revtex4}

\usepackage{amssymb}
\usepackage{graphicx}

\begin{document}

\title{On the true nature of renormalizability  in Horava-Lifshitz gravity}

\author{Fabio Briscese$^{a,b}$\footnote{E-mail: briscese.phys@gmail.com},
Yeinzon Rodr\'{\i}guez$^{a,c}$\footnote{E-mail:
yeinzon.rodriguez@uan.edu.co}, Guillermo A.
Gonz\'alez$^{a}$\footnote{E-mail: guillego@uis.edu.co}.}
\affiliation{$^{a}$ Escuela de F\'{\i}sica, Universidad Industrial
de
Santander,\\
Ciudad Universitaria, Bucaramanga 680002, Colombia.\\
$^{b}$ Istituto Nazionale di Alta Matematica Francesco Severi,\\
Gruppo Nazionale di Fisica Matematica,\\
Citt$\grave{a}$ Universitaria, c.a.p. 00185, Rome, Italy.\\
$^{c}$ Centro de Investigaciones, Universidad Antonio Nari\~no,\\
Cra 3 Este \# 47A - 15, Bogot\'a D.C. 110231, Colombia.}

\begin{abstract}
We argue that the true nature of the renormalizability of
Horava-Lifshitz gravity lies in the presence of higher order
spatial derivatives and not in the anisotropic Lifshitz scaling of
space and time. We  discuss the possibility of constructing
a higher order spatial derivatives model that has the same
renormalization properties of Horava-Lifshitz gravity but that
does not make use of the Lifshitz scaling. In addition,  the
state-of-the-art of the Lorentz symmetry restoration in
Horava-Lifshitz-type theories of gravitation is reviewed.

\end{abstract}

\maketitle

The quantum gravity problem, namely the need of an unified
description of all fundamental interactions in the same
theoretical framework, is one of the most challenging subjects of
modern physics. The most important  approaches to quantum gravity
are Loop Quantum Gravity, Strings, and Noncommutative Geometries
(see \cite{smolin} for an introductory reading), but recently a
new model of gravity,  the so called Horava-Lifshitz Gravity (H-LG)
\cite{horava}, has been proposed and it is believed to be
renormalizable. H-LG has been shown to be a viable model of
gravitation at cosmological and astrophysical levels
\cite{Mukohyama,muko2,muko3,horava2}, and black hole thermodynamics has been also
considered \cite{BH1,BH2,BH3,BH4}, but one of the main problems
with such a model is that it implies the loss of Lorentz symmetry.

H-LG makes use of two different assumptions: the introduction of
higher order spatial derivatives \footnote{With spatial
derivatives we mean covariant derivatives constructed with the
spatial components of the  metric tensor.} in the lagrangian
density of the gravitational field and the anisotropic Lifshitz
scaling of space and time.

Due to higher order spatial derivatives,   the invariance of
General Relativity (GR) under diffeomorphisms $Diff(\textit{M})$
is broken. Therefore,  the spacetime $M$ is assumed to be a
codimension-one foliable and differentiable manifold and the
theory is invariant under the group of foliation preserving
diffeomorphisms $Diff_{\textit{F}}(\textit{M})$, which leave the
spatial derivatives invariant. We stress that the invariance of
the theory under $Diff_{\textit{F}}(\textit{M})$ does not imply
the Lifshitz scaling.

Higher derivative renormalizable theories of gravitation have been
first presented in \cite{stelle} which are invariant under GR
diffeomorphysms  $Diff(\textit{M})$. The main problem here is that
they contain a physical ghost (a state of negative norm) and
therefore they violate the unitarity. Recently, a new class of
higher derivative theories has been presented that turns out to be
both renormalizable and ghost-free
\cite{moffat,tomboluis,modesto1,modesto2,modesto3}. The price to pay here is that the
theory should contain an infinite number of derivatives to avoid
the loss of unitarity. However, H-LG is naturally ghost-free
\cite{Mukohyama} since it contains higher order space derivatives
but no higher order time derivatives. Moreover, the
renormalizability condition only requires space derivatives of
finite order, therefore H-LG is simpler than an infinite order
theory. Despite of these advantages, there is an apparently strong
problem in H-LG: the loss of diffeomorphism invariance and then
the loss of Lorentz symmetry in flat spacetime as discussed
in \cite{pospelov,Bogdanos:2009uj,bemfica1,bemfica2}.

In this Letter we argue that the true nature of the
renormalizability of H-LG lies in the presence of higher order
spatial derivatives in the lagrangian density of the gravitational
field and not in the anisotropic Lifshitz scaling of space and
time.

The power counting renormalizability of H-LG is deduced by an argument based on dimensional analysis \cite{horava,Mukohyama}
which makes use of the Lifshitz scaling. As we will show, such a
dimensional analysis is incorrect and the Lifshitz scaling results
to be unnecessary for the renormalizability of the theory.
Therefore  the renormalizability of H-LG would be due uniquely  to
the presence of higher order spatial derivatives in the lagrangian
density of the gravitational field and not due to the Lifshitz scaling. In
fact, because of the higher order derivatives, the graviton propagator
goes to zero more rapidly than $1/k^2$ at high wave numbers $k$,
ensuring  the convergence of Feynman diagrams.

We stress that Lifshitz scaling and higher order spatial
derivatives are two distinct features of H-LG which are not
equivalent. As an example of this fact, we
will introduce in (\ref{Newaction}) a new gravitational model with higher order spatial
derivatives but that does not make use of the Lifshitz scaling. In
particular this model has the same gravitational lagrangian
density of H-LG and, therefore, has the same graviton propagators
and renormalization properties of H-LG. Conversely, assuming
Lifshitz scaling but no higher order derivatives, it does not lead to
a renormalizable theory since in that case the graviton
propagator scales as $1/k^2$  as in GR. These examples just
confirm the fact that the Lifshitz scaling is not necessary for the
renormalizability of H-LG-type theories of gravitation.

We remark that the new model (\ref{Newaction}) is invariant under
$Diff_{\textit{F}}(\textit{M})$ since it contains higher order
spatial derivatives that break the $Diff(\textit{M})$ invariance
of GR, but it does not use the Lifshitz scaling.  However, the breaking of
$Diff(\textit{M})$ implies the loss of Lorentz invariance. For
that reason,  we will discuss the possibility of recovering the
Lorentz symmetry in flat spacetime background in H-LG-type
theories in the Standard Model (SM) particle sector at energies
below the Planck scale.

In what follows, we will first resume the main properties of H-LG
and show  that the power counting renormalization argument
introduced in \cite{horava} is incorrect. Then,  we will argue
that the renormalization of the model  is due to the presence of
higher order spatial derivatives and not due to the Lifshitz
sclaing.  We will introduce a new higher derivative model
that has the same renormalization properties of H-LG but that does
not make use of the Lifshitz scaling. Finally we will discuss the
Lorentz symmetry restoration in Horava-Lifshitz-type theories as
known in the literature.

Differently from GR, H-LG consider space and time as fundamentally
different quantities. In fact H-LG assumes an anisotropic scaling,
or Lifshitz scaling, of space and time given by
\begin{equation}\label{anisotropicscaling}
t \rightarrow b^z \, t \,,\qquad \overrightarrow{x}\rightarrow b \,
\overrightarrow{x} \,,
\end{equation}
where $z$ is the dynamical critical exponent of the theory. Eq.
(\ref{anisotropicscaling}) is equivalent to fix the dimensions of
time and space coordinates as $[t]= [x]^z$. Defining a formal
symbol $p$ having dimension of momentum, and taking $\hbar = 1$,
one has $[t]= [p]^{-z}$ and $[x]= [p]^{-1}$ (see
\cite{visser,visser2}) and therefore one has $[t]= -z$ and $[x]=
-1$ in momentum units $[p]$. Here and below, all the dimensions
will be expressed in momentum units $[p]$. This scaling law is
believed to imply the renormalizability of the theory
\cite{horava,Mukohyama}.

The fundamental symmetry of the theory is the invariance under
$Diff_{\textit{F}}(\textit{M})$, that in a local reference frame
can be expressed as
\begin{equation}\label{hlgdiff}
t' = t'(t),\qquad \overrightarrow{x}' =
\overrightarrow{x}'(\overrightarrow{x},t) .
\end{equation}
Note that the invariance under $Diff_{\textit{F}}(\textit{M})$
does not imply the Lifshitz scaling law
(\ref{anisotropicscaling}). Moreover, one can always perform
the transformation $t' = t^{1/z}$ and $x' = x$ allowed by the
$Diff_{\textit{F}}(\textit{M})$ symmetry, which reduces to
isotropic scaling of space and time coordinates $[t']=[x']= -1$.
This gives an intuition on the fact that the Lifshitz scaling cannot be relevant for the renormalizability of the theory.

Also note that Lorentz symmetry is not a symmetry of the theory
since it is not included in $Diff_{\textit{F}}(\textit{M})$.

Due to Lifshitz scaling the speed of light now has $z$ dependent
dimensions in momentum units $[c] = [dx/dt] = z-1$
\cite{visser,visser2}. Therefore one can not simply take $[c] =1$,
but it is convenient to explicitly express the dependence on $c$
in all physical quantities. To construct the lagrangian of the
theory in D+1 dimensions, one can write the D+1-dimensional metric
tensor $g_{ij}^{(D+1)}$ in the ADM formalism \cite{adm} as
\begin{equation}\label{ADMmetric}
ds^2 = - N^2 (c \, dt)^2 + 2 \, N_\alpha \, c \, dt \, dx^\alpha +
g_{\alpha \beta}^{(D)} dx^\alpha dx^\beta \,,
\end{equation}
where Greek indices run from 1 to D and Latin indices from 0 to D,
and $N=g_{00}^{(D+1)}$, $N_\alpha=g_{0\alpha}^{(D+1)}$ and
$g_{\alpha \beta}^{(D)} = g_{\alpha \beta}^{(D+1)}$ are the lapse,
the shift and the D-dimensional spatial metric tensor
respectively. Due to the invariance under
$Diff_{\textit{F}}(\textit{M})$, the lapse $N(t)$ is required to
be a function of time only.

Note that the GR time variable  $x^0 = c t$  has dimensions
$[x^0]=-1$ in momentum units. The GR prescription implies the use of the
variable  $x^0$ instead of $t$ in all physical quantities. Of
course one can always write the line element as $ds^2 = - N^2 dt^2
+ 2 \, N_\alpha \, dt \, dx^\alpha + g_{\alpha \beta}^{(D)}
dx^\alpha dx^\beta$. However, in such a case, the dimension of $c$
would be absorbed in the dimensions of the lapse and shift and one
would have $[N] = [N_\alpha] = z-1$ so the physics would remain
unchanged.

The action of the theory is \cite{Mukohyama}
\begin{equation}\label{HLaction}
S_{H-LG} \equiv \frac{2}{k^2} \int  c\, dt \, d^Dx \, N
\sqrt{-g^{(D)}} \left( \textit{L}_{H-LG} + \textit{L}_{SM} \right) \,,
\end{equation}
where $\textit{L}_{H-LG}$ is the H-LG lagrangian density and
$\textit{L}_{SM}$ is the usual SM lagrangian density.
$\textit{L}_{H-LG}$ contains the higher order spatial derivatives
that should ensure the renormalizability of the theory. Its
explicit form is given by \cite{Mukohyama}

\begin{equation}\label{HL Lagraingian}
\textit{L}_{H-LG} \equiv K^{\alpha \beta}K_{\alpha \beta}- \lambda
K^2 + R^{(D)} -2 \Lambda + I_{z=2} + I_{z=3} \,,
\end{equation}
where $\lambda$ and $\Lambda$ are constants, $R^{(D)}$ is the
Ricci scalar constructed with $g_{\alpha \beta}^{(D)}$, $K_{\alpha
\beta} $ is the extrinsic curvature defined as
\begin{equation}\label{Kextrinsic curvature}
K_{\alpha \beta} \equiv \frac{1}{2N} \left(\partial_{ct}
g^{D}_{\alpha \beta} - D_\alpha N_\beta - D_\beta N_\alpha
\right),
\end{equation}
$k \equiv K^\alpha_\alpha$, $D_\alpha$ is the covariant derivative
constructed with $g^{(D)}_{\alpha \beta}$ and
\begin{eqnarray}
\label{Iz} I_{z=3} &\equiv& c_1 D_\alpha R^{(D)}_{\beta \gamma}
D^\alpha R^{(D) \,\beta \gamma} + c_2 D_\alpha R^{(D)} D^\alpha
R^{(D)} + \nonumber \\
&& +c_3 R^{(D) \,\alpha}_\beta R^{(D) \,\gamma}_\alpha R^{(D)
\,\beta}_\gamma
+ c_4 R^{(D)} \, R^{(D) \,\alpha}_\beta R^{(D) \,\beta}_\alpha + \nonumber \\
&& + c_5 R^{(D) \,3} \,,\\
&& \nonumber \\
I_{z=2} &\equiv& c_6 R_\alpha^{(D) \,\beta} R^{(D) \,\alpha}_\beta +
c_7 R^{(D) \,2}.
\end{eqnarray}
H-LG has a scalar  $\zeta$ and a transverse traceless tensor
$h_{\alpha \beta}$ physical degrees of freedom defined by
$g_{\alpha \beta} = (1 + 2 \, \zeta) \delta_{\alpha \beta} +
h_{\alpha \beta}$, whose dispersion relations are
\begin{eqnarray}\label{disperion relations H-LG}
\omega_\zeta^2 = \frac{\lambda-1}{3\lambda-1}\left(
\frac{k^6}{M_s^4}+ \kappa_s \frac{k^4}{M_s^2}-k^2
\right) \,, \\
&& \nonumber \\
\omega_h^2 = \frac{k^6}{M_t^4}+ \kappa_t
\frac{k^4}{M_t^2}+k^2 \,, \label{drHLG2}
\end{eqnarray}
where \cite{Mukohyama}
\begin{eqnarray}
M_s^{-4} &\equiv& -2 ( 3 \, c_1 + 8 \, c_2) M_P^{-2} \,, \\
\kappa_s \, M_s^{-2} &\equiv& -2 ( 3 \, c_6 + 8 \, c_7) M_P^{-2} \,, \\
M_t^{-4} &\equiv& - 2 \, c_1 \, M_P^{-2} \,, \\
\kappa_t M_t^{-2} &\equiv& - 2 \, c_6 \, M_P^{-2} \,,
\end{eqnarray}
and $M_P$ is the reduced Planck mass.
We mention that the scalar degree of freedom $\zeta$ does not
decouple in the low energy limit but it can be eliminated by
requiring local scale invariance as in shape dynamics \cite{shape
dynamics}.

Comparing Eq. (\ref{HLaction}) with the Einstein-Hilbert action of
GR, we can see that the GR diffeomorphism
invariance $Diff(\textit{M})$ is broken at two different levels.
First, it is broken at the level of the density lagrangian
$\textit{L}_{H-LG}$, since this is constructed with quantities
explicitly invariant under $Diff_{\textit{F}}(\textit{M})$;
second, it is broken at the level of the D+1 spacetime measure $c
\,dt d^Dx N \sqrt{-g^D}$.

In Ref. \cite{horava}, the assumption of the Lifshitz scaling law
(\ref{anisotropicscaling}) is motivated since it would lead to the
power counting renormalizability of the theory. This conclusion is
based on the following dimensional argument. First of all note
that in \cite{horava} the author uses Eqs.
(\ref{ADMmetric})-(\ref{HL Lagraingian}) with $c=1$ ignoring the
speed of light $c$ and its dimensions. This implies that the
dimensions of $c$ are included in the dimensions of other physical
quantities. For example in \cite{horava} the author uses $[N]=0$,
$[N_\alpha] = z-1$ (Eq. (2.4)) and $[K_{ij}] = z$. By means of
such a dimensional analysis, and requiring the dimensionless of the
action (\ref{HLaction}), one would conclude that the dimension of
the constant $k$ is $(z-D)/2$ (Eq. (21) in \cite{horava}). This
would lead to the conclusion that the dimension of $k$  would be
non negative for $ z \geq D$ and that, therefore, the theory would
be power counting renormalizable.

Unfortunately, this dimensional argument is not correct and this
is evident if one includes explicitly the speed of light $c$ in
Eqs. (\ref{ADMmetric})-(\ref{HL Lagraingian}) as we did. From Eq.
(\ref{ADMmetric}), it is immediate to recognize that the lapse and
the shift should have the same dimension $[N] = [N_\alpha] = 0$
since $[ds^2] = -2$. This also implies that $[L_{H-LG}] = 2
[K_{ij}] = 2$. Requiring the dimensionless of the action
$[S_{H-LG}]=0$ and using Eq.(\ref{HLaction}) with the correct $c
\, dt$ factor in the integral, one concludes that the constant $k$
has dimension $[k]=(1-D)/2$ independently of $z$ as it should be.
We stress again that the difference between the two dimensional
analysis lies on the fact that in Ref. \cite{horava} the
dependence on $c$ is not considered explicitly. We also stress
that, as noted before, even if one does not consider explicitly
the light speed $c$, the dimensions of the lapse and shift
functions should be the same $[N]=[N_\alpha]=z-1$ and, therefore,
Eq. (2.4) of \cite{horava} is not correct. Using the correct
dimensions $z-1$ of the lapse and shift functions, one again
obtains the correct dimension $[k]=(1-D)/2$ which is independent of $z$.

Based on such observations, one can conclude that the
renormalizability of H-LG cannot be motivated by the dimensional
analysis performed in \cite{horava}. Whereas the power counting
argument used in \cite{horava} has been shown to be incorrect, the
renormalizability of H-LG remains to be demonstrated rigorously.
We notice that in \cite{orlando}, the renormalizability of H-LG is
related to that of topologically massive gravity but, as the
authors remark, the latter is not strictly proven.

Therefore, we conclude that the Lifshitz scaling has no any role in the
renormalizability of H-LG since the correct dimensional analysis
has been shown to be independent of the dynamical critical
exponent $z$. We think that the true nature of the
renormalizability of H-LG lies in the presence of higher
derivative terms in the lagrangian and not in the scaling law
(\ref{anisotropicscaling}).

In fact, renormalizability seems to be a feature of higher
derivative theories \cite{stelle,moffat,tomboluis,modesto1,modesto2,modesto3}, due to
the fact that the presence of higher order derivatives implies
that the graviton propagator goes to zero more rapidly than
$1/k^2$  at high wave numbers $k$. However, in order to be
unitary, such higher derivative theories should contain an
infinite number of derivatives \cite{moffat,tomboluis,modesto1,modesto2,modesto3}.
The advantage of H-LG with respect to higher order derivative
theories is that it is naturally ghost-free \cite{Mukohyama} and
includes a limited number of space derivatives but with the
handicap of losing Lorentz symmetry in flat spacetime.

Let us introduce a new model with higher spatial derivatives
that has the same renormalization properties of H-LG but that does
not make use of the Lifshitz scaling law
(\ref{anisotropicscaling}). This example explicitly shows  that
the Lifshitz scaling is unnecessary for the renormalizability.
Moreover, it can can be of interest since, as we will see, it is
more conservative than H-LG since it does not make use of the
Lifshitz scaling and uses the same volume element of GR. This new
model is only introduced here but its main properties deserve
further study.

Let us use the usual scaling of time and space $[t]=[x]=-1$. This
legitimates the  use of natural units with $c=1$. Moreover, in
order to be conservative, we can break the diffeomorphism
invariance of GR only at the level of the gravitational lagrangian
density and take the same D+1 dimensional measure and SM
lagrangian density as in GR.  We restrict to the case D=3 and
write the action for such a new theory as
\begin{equation}\label{Newaction}
S \equiv \frac{2}{k^2} \int d^4x \, \sqrt{-g^{(3+1)}} \left(
\textit{L}_{H-LG} + \textit{L}_{SM} \right) \,,
\end{equation}
where $\textit{L}_{H-LG}$ is the same as in (\ref{HL Lagraingian})
and   $d^4x = dt d^3x$.  Taking $\lambda = 1$, $k^2 = 32 \pi G$,
and $\Lambda=0$, the action (\ref{Newaction}) describes GR plus
higher order derivatives \cite{Mukohyama}. For simplicity, we use
the same lagrangian density $\textit{L}_{H-LG}$ as H-LG since it
contains higher order spatial derivatives that should ensure the
renormalizability of the theory, but it is clear that one can make
a different choice, e.g. adding
$Diff_{\textit{F}}(\textit{M})$-invariant terms containing even
higher order spatial derivatives. Also note that (\ref{Newaction})
reduces to (\ref{HLaction}) in the synchronous gauge when $D=3$.

Since in Eq. (\ref{Newaction}) we take the same lagrangian density
$L_{H-LG}$ of H-LG, expanding the gravitational field around a
flat spacetime background gives the same kinetic term of H-LG.
Therefore, one has the same scalar $\zeta$ and traceless tensor
$h_{\alpha \beta}$ gravitational degrees of freedom and the same
dispersion relations (\ref{disperion relations H-LG}) and
(\ref{drHLG2}) as in H-LG and then the same propagators for the
$\zeta$ and $h_{\alpha \beta}$ fields. This implies that
(\ref{Newaction}) has the same renormalization properties of H-LG.
However the renormalizability of (\ref{Newaction}), as well as that
of H-LG, remains to be proved rigorously.

We remark that  in the model (\ref{Newaction}), the parameter $z=3$
is no longer the dynamical critical exponent of the theory (which
does not assume Lifshitz scaling) but is simply related to the
order $2 \, z = 6$ of spatial derivatives. Therefore, the UV
behavior $\sim k^6$ of the graviton propagators is due to higher
order derivatives and is not related to the Lifshitz scaling.

Of course, the new quantum gravity framework described by
(\ref{Newaction})  should be tested in the pure gravitational
sector, meaning that one should study the behavior of gravity at
cosmological and astrophysical levels.  These topics deserve
further study.

We remark as well that in this new theory one has to deal with the same
SM Lorentz restoration problem of H-LG \cite{pospelov}. Therefore,
here we discuss Lorentz symmetry restoration in H-LG-type
gravities.

It is evident from (\ref{HLaction}) that, if the gravitational
field is turned off, the residual SM action is Lorentz invariant.
Let us consider SM particles in a flat background spacetime plus
quantum gravitational perturbations. Note that SM particle
propagators are Lorentz invariant at tree level, but graviton
propagators are not since the lagrangian density of the
gravitational field is explicitly not a Lorentz invariant. Of
course, loop corrections will contain graviton propagators and,
therefore, will break Lorentz invariance of SM particle
propagators.

Since in (\ref{HLaction}) SM particles are minimally coupled to
gravity, this implies that, as in GR, interactions between SM
particles and gravitons are expected to be Planck scale
suppressed. Therefore one expects that Lorentz symmetry breaking
loop corrections to SM propagators  are Planck scale suppressed
and Lorentz symmetry is safe.

Unfortunately, this is not the case since, as shown in
Ref. \cite{pospelov}, one loop Lorentz symmetry breaking effects are
Planck scale suppressed but coupled to quadratically divergent
diagrams and  results to be $\propto (\Lambda_{HL}/M_P)^2
\Lambda_{UV}^2$ where $\Lambda_{HL} \sim c_1,c_2,\ldots, c_5$, is
the Lorentz symmetry breaking scale and $\Lambda_{UV}$ is the
ultraviolet (UV) cutoff. That means that Lorentz symmetry
breaking in the SM sector is never negligible in this picture. In
Ref. \cite{pospelov}, the authors propose to introduce higher
derivative terms as $K_{\alpha \beta} \Delta K^{\alpha \beta}$ or
$\nabla_\alpha K^{\alpha \beta} \nabla^\gamma K^{\gamma}_{\beta}$
in order to regularize such a quadratically divergence, but this
possibility has not been yet completely explored.

Lorentz symmetry restoration is a hard problem for H-LG-type
models. In fact, if Lorentz symmetry is broken in the SM sector
below the Planck scale, the SM particle propagators would not be
Lorentz invariant and the Minkowskian energy-momentum
``dispersion'' relation would be deformed below the Planck scale.
Such dispersion relation deformations have been shown to be
testable in some context up to the Planck scale by using
astrophysical data
\cite{charmousis,iengo,mukohyama2,xiao,Shao,astrophysicalregime,a1,a2,a4,a5,fermispacetelescope,f1,f2,f3,f4,amelino
horava} or in cold atoms recoil frequency experiments
\cite{coldatoms1,coldatoms2} and, therefore, a sub-Planckian
Lorentz symmetry breaking  could rule out the model.

The scope of this Letter is to generate a debate about the true
nature of the renormalizability of H-LG. What is important here is
to show that the anisotropic scaling law
(\ref{anisotropicscaling}) is not necessary to ensure the
renormalization of H-LG. We argued that such renormalizability
would be due only to higher order derivatives and not due to the Lifshitz
scaling (\ref{anisotropicscaling}). We have  introduced a new
higher spatial derivatives  model that does not make use of the
Lifshitz scaling an that does not break $Diff(\textit{M})$ in the
spacetime measure. This model  results more conservative but  has
the same renormalization properties of H-LG. Finally, we have
discussed the Lorentz invariance restoration problem in H-LG-type
theories as known in the literature.

\textit{Acknowledgments}: We thank A. Marcian\'{o}, D. H. Lyth, S.
Mukohyama, M. Sasaki, and M. de Llano for useful discussions during
the edition of this Letter.  F.B. is a UIS postdoctoral fellow. Y.R. is supported by DIEF de Ciencias (UIS) grant number 5177.


\begin{thebibliography}{99}


\bibitem{smolin} L. Smolin, {\it Three roads to quantum gravity },
London, UK: Weidenfeld \& Nicolson (2000).

\bibitem{horava} P. Ho\v{r}ava, Phys. Rev. D {\bf 79}, 084008 (2009).

\bibitem{Mukohyama} S. Mukohyama, Class. Quantum Grav. {\bf 27}, 223101 (2010).

\bibitem{muko2} K. Izumi and S. Mukohyama, Phys. Rev. D {\bf 84}, 064025 (2011).

\bibitem{muko3} A. E. Gumrukcuoglu, S. Mukohyama, and A. Wang, Phys. Rev. D {\bf 85}, 064042 (2012).

\bibitem{horava2} P. Ho\v{r}ava and C. M. Melby-Thompson, Phys. Rev. D {\bf 82}, 064027 (2010).

\bibitem{BH1} R. Biswas and S. Chakraborty,  Gen. Rel. Grav. {\bf 43} 41
(2011).

\bibitem{BH2} W. Janke, D. A. Johnston, and R. Kenna,     J. Phys. A {\bf 43}, 425206 (2010).

\bibitem{BH3} Q.-J. Cao, Y.-X. Chen, and K.-N. Shao, Phys. Rev. D {\bf 83}, 064015 (2011).

\bibitem{BH4} H. Quevedo, A. S\'anchez, S. Taj, and A. V\'azquez,  J. Phys. A {\bf 45}, 055211 (2012).

\bibitem{stelle} K. S. Stelle, Phys. Rev. D {\bf 16}, 953 (1977).

\bibitem{moffat} J. W. Moffat, Eur. Phys. J. Plus {\bf 126}, 43 (2011).

\bibitem{tomboluis} E. T. Tomboulis, arXiv:hep-th/9702146.


\bibitem{modesto1} L. Modesto,  arXiv:1107.2403 [hep-th].

\bibitem{modesto2} L. Modesto,  arXiv:1206.2648 [hep-th]. 

\bibitem{modesto3} L. Modesto,  arXiv:1202.3151 [hep-th].

\bibitem{pospelov} M. Pospelov and Y. Shang, Phys. Rev. D {\bf 85}, 105001 (2012).


\bibitem{Bogdanos:2009uj} C. Bogdanos and E. N. Saridakis,  Class. Quantum Grav. {\bf
27}, 075005 (2010).

\bibitem{bemfica1} F.S. Bemfica and M. Gomes, Phys. Rev. D {\bf 84}, 084022 (2011).

\bibitem{bemfica2} F.S. Bemfica and M. Gomes, arXiv:1111.5779 [hep-th].



\bibitem{visser} M. Visser, Phys. Rev. D {\bf 80}, 025011 (2009).

\bibitem{visser2} S. Weinfurtner, T.P. Sotiriou, and M. Visser, J. Phys. Conf. Ser. {\bf 222}, 012054 (2010).

\bibitem{adm} R. Arnowitt, S. Deser, and C. W. Misner, {\it Gravitation: an introduction to current research}, New York, USA: Wiley (1962).

\bibitem{shape dynamics} H. Gomes, S. Gryb, and T. Koslowski, Class. Quantum Grav. {\bf 28}, 045005 (2011).


\bibitem{orlando} D. Orlando and S. Reffert, Class. Quantum Grav. {\bf 26}, 155021 (2009).


\bibitem{charmousis} C. Charmousis, G. Niz,
A. Padilla, and P. M. Saffin, JHEP {\bf 0908}, 070 (2009).

\bibitem{iengo} R. Iengo, J. G. Russo, and M. Serone, JHEP {\bf 0911}, 020 (2009).


\bibitem{mukohyama2} S. Mukohyama, K. Nakayama, F. Takahashi,  and S. Yokoyama, Phys. Lett. B {\bf 679}, 6 (2009).

\bibitem{xiao} Z. Xiao and B.-Q. Ma, Phys. Rev. D {\bf 80}, 116005 (2009).

\bibitem{Shao} L. Shao, Z. Xiao, and B.-Q. Ma, Astropart. Phys. {\bf 33}, 312 (2010).

\bibitem{astrophysicalregime} G. Amelino-Camelia, {\it et. al.}, Nature \textbf{393}, 763 (1998).



\bibitem{a1} B. E. Schaefer, Phys. Rev. Lett. \textbf{82}, 4964 (1999).

\bibitem{a2} S. D. Biller \textit{et. al.}, Phys. Rev. Lett. \textbf{83}, 2108 (1999).

\bibitem{a4} T. Kifune, Astrophys. J. Lett. \textbf{518}, L21 (1999).

\bibitem{a5} G. Amelino-Camelia, Nature \textbf{408}, 661 (2000).

\bibitem{fermispacetelescope} A. A. Abdo \textit{et. al.},
Science \textbf{323}, 1688 (2009).

\bibitem{f1} J. Ellis, N. E. Mavromatos, and D. V. Nanopoulos, Phys. Lett. B
\textbf{674}, 83 (2009).

\bibitem{f2} G. Amelino-Camelia and L. Smolin, Phys. Rev. D \textbf{80},
084017 (2009).

\bibitem{f3} A. A. Abdo \textit{et al.}, Nature \textbf{462}, 331 (2009).

\bibitem{f4} G. Amelino-Camelia, Nature \textbf{462}, 291 (2009).


\bibitem{amelino horava}G. Amelino-Camelia, L. Gualtieri, and F. Mercati, Phys. Lett. B {\bf 686}, 283 (2010).



\bibitem{coldatoms1} G. Amelino-Camelia, C. L\"{a}mmerzahl, F. Mercati, and
G. M. Tino, Phys. Rev. Lett. \textbf{103}, 171302 (2009).

\bibitem{coldatoms2} F. Mercati {\it et. al.}, Class. Quantum Grav. \textbf{27}, 215003 (2010).


\end{thebibliography}
\end{document}